# Mutual Events in the Cold Classical Transneptunian Binary System Sila and Nunam

W.M. Grundy[a], S.D. Benecchi[b], D.L. Rabinowitz[c], S.B. Porter[a],
L.H. Wasserman[a], B.A. Skiff[a], K.S. Noll[d], A.J. Verbiscer[e],
M.W. Buie[f], S.W. Tourtellotte[g], D.C. Stephens[h], and H.F. Levison[f]

a. Lowell Observatory, Flagstaff AZ 86001.
b. Carnegie Institution of Washington, Department of Terrestrial Magnetism, Washington DC 20015.
c. Center for Astronomy and Astrophysics, Yale University, New Haven CT 06520.
d. NASA Goddard Space Flight Center, Greenbelt MD 20771.
e. Department of Astronomy, University of Virginia, Charlottesville VA 22904.
f. Southwest Research Institute, Boulder CO 80302.
g. Astronomy Department, Yale University, New Haven CT 06520.
h. Department of Physics and Astronomy, Brigham Young University, Provo UT 84602.

**ABSTRACT**

    Hubble Space Telescope observations between 2001 and 2010 resolved the binary components of the Cold Classical transneptunian object (79360) Sila-Nunam (provisionally designated 1997 $CS_{29}$). From these observations we have determined the circular, retrograde mutual orbit of Nunam relative to Sila with a period of 12.50995 ± 0.00036 days and a semimajor axis of 2777 ± 19 km. A multi-year season of mutual events, in which the two near-equal brightness bodies alternate in passing in front of one another as seen from Earth, is in progress right now, and on 2011 Feb. 1 UT, one such event was observed from two different telescopes. The mutual event season offers a rich opportunity to learn much more about this barely-resolvable binary system, potentially including component sizes, colors, shapes, and albedo patterns. The low eccentricity of the orbit and a photometric lightcurve that appears to coincide with the orbital period are consistent with a system that is tidally locked and synchronized, like the Pluto-Charon system. The orbital period and semimajor axis imply a system mass of (10.84 ± 0.22) × $10^{18}$ kg, which can be combined with a size estimate based on Spitzer and Herschel thermal infrared observations to infer an average bulk density of $0.72^{+0.37}_{-0.23}$ g $cm^{-3}$, comparable to the very low bulk densities estimated for small transneptunian binaries of other dynamical classes.

# 1. Introduction

Observation of mutual eclipses and occultations between components of a binary system is a powerful technique for remote characterization of small and distant objects. Mutual events have been used to constrain binary asteroid mutual orbits, shapes, and densities (e.g., Descamps et al. 2007), to monitor volcanic activity on the jovian satellite Io (e.g., Rathbun and Spencer 2010), and to distinguish surface compositions and map albedo patterns on Pluto and Charon (e.g., Binzel and Hubbard 1997). It would be valuable if mutual event observation techniques could be brought to bear on many more transneptunian objects (TNOs), since their remote locations in the Kuiper belt and their small sizes make them particularly challenging to investigate using other observational techniques. Already, mutual events have been observed in the contact (or near-contact) binary system 139775 (Sheppard and Jewitt 2004; Lacerda 2011) and in the Haumea triple system (Ragozzine and Brown 2010; although in that system the large contrast between Haumea and satellite sizes and their non-tidally locked spin states greatly complicates interpretation of mutual event data). With more and more transneptunian binaries (TNBs) being discovered (e.g., Noll et al. 2008a), the likelihood grows for additional TNB mutual events in the near future, but observations are unlikely without advance knowledge of their mutual orbits. Planning for mutual event observations is one of many motivations for the ongoing campaign of TNB orbit determination from which this paper arises (see http://www.lowell.edu/~grundy/tnbs).

Most known TNBs inhabit the "Classical" sub-population of TNOs orbiting the Sun on relatively low-inclination, low-eccentricity orbits not in mean-motion resonance with Neptune (Elliot et al. 2005; Gladman et al. 2008). Although they are less dynamically excited than other TNO orbits, Classical TNO orbits have been further subdivided into dynamically "Hot" and "Cold" Classical subgroups based on the inclinations of their heliocentric orbits (e.g., Brown 2001; Gulbis et al. 2010; although Peixinho 2008 argues that the Tisserand parameter could be a better criterion). Binaries are particularly abundant among the low inclination Classical TNOs, or "Cold Classicals" (Stephens and Noll 2006; Noll et al. 2008b). These objects are of particular interest for having accreted relatively far from the young Sun, perhaps near their present-day locations with semimajor axes in the 42 to 47 AU range. This contrasts with other dynamical classes of TNOs populated by objects thought to have formed much closer to the Sun prior to emplacement into their current orbits by dramatic events early in solar system history (e.g., Levison et al. 2008). Cold Classicals are thus seen as offering a window into conditions in the outermost parts of the nebular disk. The Cold Classical sub-population has itself recently been divided into "kernel" and "stirred" sub-components (Petit et al. 2011), where the "kernel" is a concentration of Cold Classical objects with semimajor axes near 44 AU and eccentricities near 0.06. The significance of this clump and its relation to circumstances in the protoplanetary nebula and/or subsequent dynamical erosion of the Kuiper belt is not yet clear.

In addition to their high rate of binarity, other intrinsic properties of Cold Classical TNOs also appear to be distinctive. Their distribution of colors in reflected sunlight looks more uniformly red than the broader mix of colors seen among other dynamical classes (e.g., Trujillo and Brown 2002; Tegler et al. 2003; Gulbis et al. 2006; Peixinho et al. 2008). Many of the Cold Classical binaries consist of near-equal brightness components, unlike the more asymmetric pairings seen elsewhere (Noll et al. 2008b). Their magnitude frequency distribution is much steeper than that of more excited TNOs (Bernstein et al. 2004; Fuentes and Holman 2008; Fraser et al. 2010). This distribution is often taken as a proxy for their size frequency distribution, although without knowledge of albedos, the absolute normalization between brightness and size is uncertain. Indeed, albedos reported for Cold Classical TNOs are higher than is typical of small TNOs on more excited orbits (Grundy et al. 2005a; Stansberry et al. 2008; Brucker et al. 2009; Vilenius et al. 2012). However, Cold Classicals tend toward the faint limit of what can be investigated with available observational techniques for estimating albedos. If this population actually had a broad distribution of albedos, the small sample of them studied thus far would likely be biased in favor of higher albedo objects (e.g., Parker et al. 2011). More work is needed to resolve this issue, and also to investigate whether the distinctive properties of Cold Classicals are features of just the "kernel" or "stirred" sub-components, or are shared among both. About a fifth of known Cold Classical binaries fall into the Petit et al. (2011) "kernel" region of orbital element space, roughly on par with the ratio of "kernel" to all Cold Classicals, so at least this characteristic does not seem to be confined to one or the other subgroup.

Based on its heliocentric orbital elements, the Sila and Nunam system is probably a member of the Cold Classical group. We say "probably" because the Cold and Hot Classical sub-populations overlap in orbital element space. A low inclination heliocentric orbit is required for membership in the Cold Classical group, but does not exclude membership of the Hot population. From debiased Deep Ecliptic Survey (DES) observations (Eliot et al. 2005), Gulbis et al. (2010) described the separate inclination distributions of the Hot and Cold Classical groups, enabling the probability of Cold Classical membership to be estimated as a function of inclination, as shown in Fig. 1. For the system's mean inclination $<i_\odot> = 3.84°$ (relative to the invariable plane, and averaged over a 10 Myr integration), this translates into a 76% probability of Cold Classical membership. This calculation neglects the different brightness distributions of Hot and Cold groups (e.g., Petit et al. 2011), but the brightness of the Sila and Nunam system is typical of the DES sample, so errors from this source are not expected to be large. Increasing confidence that it belongs to the Cold Classical group are its identification as a binary with near-equal brightness components (Stephens and Noll 2006). If binary probabilities for Hot and Cold Classicals are taken as 2.9% and 29%, respectively (Noll et al. 2008b), then the fact of Sila and Nunam's binarity boosts the system's probability of belonging to the Cold Classical group to 97%, via Bayes' Theorem. Considera-tion of the system's red coloration (Barucci et al. 2000) would increase the odds of Cold Classical membership still further. At 43.9 AU, Sila and Nunam's mean heliocentric semimajor axis $<a_\odot>$ is consistent with the Petit et al. (2011) "kernel" group, but its mean eccentricity $<e_\odot> = 0.013$ is lower than the 0.03 to 0.08 eccentricity range of the Petit et al. model kernel cluster, suggesting membership in the "stirred" Cold Classical group.

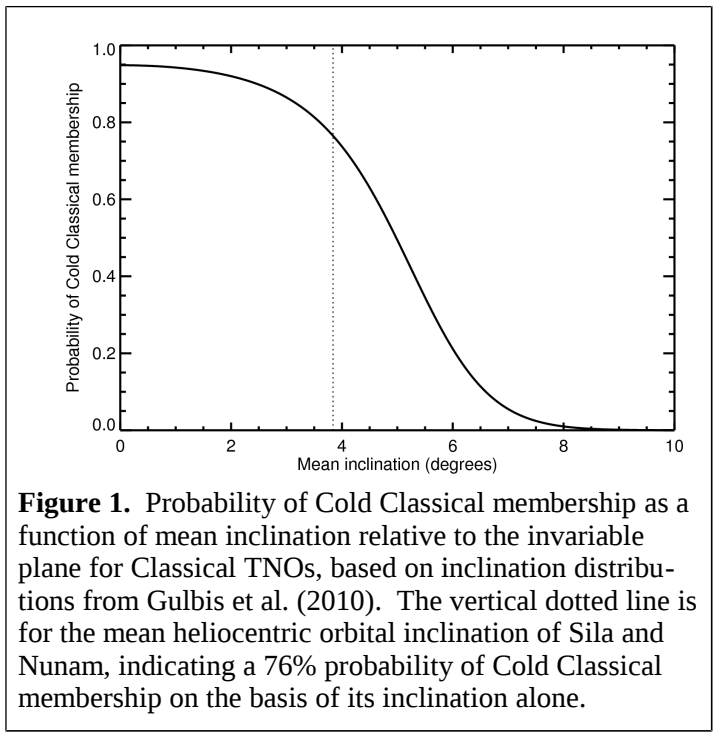

**Figure 1.** Probability of Cold Classical membership as a function of mean inclination relative to the invariable plane for Classical TNOs, based on inclination distributions from Gulbis et al. (2010). The vertical dotted line is for the mean heliocentric orbital inclination of Sila and Nunam, indicating a 76% probability of Cold Classical membership on the basis of its inclination alone.

Being among the brightest of the probable Cold Classical TNOs, the Sila and Nunam system has been targeted for more detailed study by many groups, using a variety of observational techniques. For instance, Grundy et al. (2005b) obtained a low resolution near-infrared spectrum at the Keck telescope, showing an absence of deep ice absorption bands. Photometric observa-tions by Rabinowitz et al. (2009) and Verbiscer et al. (2010) revealed that the system exhibits a narrow opposition spike. Stansberry et al. (2008), Müller et al. (2010), and Vilenius et al. (2012) reported thermal infrared observations from Spitzer and Herschel space observatories, constraining the size and pointing to a visual albedo in the range of 0.06 to 0.10. This paper reports on the determination of the mutual orbit of Sila and Nunam along with additional information that can be learned from knowledge of this orbit and from observing the mutual events it produces.

## 2. Astrometric Observations

The Hubble Space Telescope (HST) acquired images of Sila and Nunam through five different programs, each using a different instrument: 9110 (STIS), 9386 (NICMOS/NIC2), 10514 (ACS/HRC), 11178 (WFPC2/PC), and 11650 (WFC3/UVIS). We measured relative astrometry of the two components from these observations by fitting a pair of point-spread-functions (PSFs) generated by Tiny Tim (Krist and Hook 2004) to the two components in each image. Astrometric uncertainties were estimated from the scatter of PSF-fits to a series of dithered frames obtained during each HST visit to the system (except for the program 9110 STIS observations, where only a single frame was acquired during each visit). Details of these procedures have been published previously, and in the interest of brevity we refer interested readers to those papers (e.g., Grundy et al. 2009, 2011). Table 1 lists our astrometric measurements and estimated 1-σ uncertainties for Nunam's position relative to Sila. So, for instance, on 2007 October 9 at a mean time of 21:22 UT, Nunam appeared 20.6 ± 1.5 mas West and 85.3 ± 1.4 mas North of Sila, and was fainter by 0.10 ± 0.05 magnitudes.

**Table 1.**
Astrometric Data From Hubble Space Telescope.

| Mean UT observation date and hour | Instrument[a] | $r$[b] | $\Delta$[b] | $g$[b] | $\Delta x$[c] | $\Delta y$[c] | $V_{Sila}$ | $V_{Nunam}$ |
|---|---|---|---|---|---|---|---|---|
| | | (AU) | | (°) | (arcsec) | | (magnitudes)[d] | |
| 2001/11/01  2.2502 | STIS | 43.580 | 43.441 | 1.29 | +0.025(10) | −0.065(10) | - | - |
| 2001/11/04 13.6999 | STIS | 43.579 | 43.382 | 1.28 | −0.023(10) | +0.082(10) | - | - |
| 2002/10/22 23.6723 | NICMOS | 43.572 | 43.617 | 1.31 | −0.0303(44) | +0.0680(21) | - | - |
| 2005/11/30 12.4176 | ACS | 43.544 | 43.012 | 1.10 | +0.0064(19) | +0.0118(12) | - | - |
| 2007/10/09 21.3686 | WFPC2 | 43.525 | 43.900 | 1.21 | −0.0206(15) | +0.0853(14) | 22.83±0.03 | 22.93±0.04 |
| 2007/10/10  2.2353 | WFPC2 | 43.525 | 43.896 | 1.22 | −0.0226(24) | +0.0842(13) | 22.82±0.03 | 22.91±0.04 |
| 2007/10/11 16.5311 | WFPC2 | 43.525 | 43.871 | 1.23 | −0.0199(11) | +0.0493(17) | 22.78±0.03 | 23.13±0.04 |
| 2007/11/22 10.5520 | WFPC2 | 43.524 | 43.164 | 1.22 | +0.0252(10) | −0.0820(16) | 22.86±0.05 | 22.89±0.02 |
| 2007/12/23 13.9242 | WFPC2 | 43.523 | 42.737 | 0.78 | −0.0230(10) | +0.0851(10) | 22.70±0.03 | 22.88±0.02 |
| 2010/02/27 17.1597 | WFC3 | 43.507 | 42.638 | 0.63 | +0.0134(44) | −0.0255(25) | - | - |
| 2010/02/28 16.7486 | WFC3 | 43.507 | 42.646 | 0.65 | +0.009(10) | +0.0072(50) | - | - |
| 2010/04/03 16.0096 | WFC3 | 43.506 | 43.062 | 1.19 | +0.0237(14) | −0.0864(13) | 22.50±0.04 | 22.44±0.04 |
| 2010/04/26 18.1613 | WFC3 | 43.506 | 43.440 | 1.32 | +0.0046(38) | −0.0455(22) | 22.57±0.03 | 22.56±0.03 |
| 2010/05/28 18.7658 | WFC3 | 43.505 | 43.965 | 1.18 | −0.0082(32) | +0.0677(25) | 22.50±0.04 | 22.59±0.04 |

Table notes:

[a.] The camera used with NICMOS was NIC2, the camera used with ACS was the HRC, the camera used with WFPC2 was the PC, and the camera used with WFC3 was UVIS.

[b.] The distance from the Sun to the target is $r$ and from the observer to the target is $\Delta$. The phase angle, the angular separation between the observer and Sun as seen from the target, is $g$.

[c.] Relative right ascension $\Delta x$ and relative declination $\Delta y$ are computed as $\Delta x = (\alpha_2 - \alpha_1)\cos(\delta_1)$ and

$\Delta y = \delta_2 - \delta_1$, where α is right ascension, $\delta$ is declination, and subscripts 1 and 2 refer to Sila and Nunam, respectively. Estimated 1-σ uncertainties in the final 2 digits are indicated in parentheses. Uncertainties are estimated from the scatter between fits to the individual frames of a dithered series, except for STIS observations where only a single frame was taken.

d. Separate *V* filter photometry for Sila and Nunam is reported, where available. Visits lacking *V* equivalent observations, or having insufficient spatial separation to extract separate photometry are indicated with dashes. Examples of the former include STIS and ACS observations taken through a clear filter, and NICMOS observations done at near-infrared wavelengths. The WFC3 observations on 2010 February 27 and 28 are cases where light from Sila and Nunam was too blended to permit separate photometry.

## 3. Orbit Solution

Keplerian orbits for Nunam's motion relative to Sila were fitted to the relative astrometric data following methods described earlier (e.g., Grundy et al. 2009, 2011). The near-equal brightnesses of Sila and Nunam complicated this task. For observations separated by a long time interval, there was no reliable way to identify which of the two components at one epoch corresponded to each component at the other epoch, prior to knowing the orbit. To overcome this difficulty, we created a single bit identity variable for each observation except for the three executed in rapid succession in 2007 October, where we could be reasonably confident that little motion had occurred between successive visits. After each new observation, we used Monte Carlo techniques to assess the probability density function (PDF) in orbital element space, allowing all possible permutations of the identity bits during the generation of the random orbits (see Grundy et al. 2008 for details). The number of possible permutations of these identity bits is $2^n$ where $n$ is the number of observations having ambiguous identities. To give an idea of the computational cost of considering these permutations, after the 12$^{th}$ observation, the identities were uncertain in all but the three mentioned earlier, so $2^{(12-3)}$ or 512 permutations had to be considered, of which only 6 actually contributed appreciably to the PDF. After the 13$^{th}$ visit, this was up to 1024 permutations, although only 2 still contributed meaningfully to the PDF. The Monte Carlo collection of orbits was reprojected to the sky plane as a function of time to identify optimal follow-up times for excluding incorrect permutations of the identity bits. To satisfy ourselves that a unique solution had finally been found, we required all but one of the multiple solutions permitted by the various permutations to be excluded at 99% confidence, which did not happen until observations were obtained at a total of 14 separate epochs, resulting in the identities listed in Table 1. Our assignments of "primary" and "secondary" are somewhat arbitrary in this system where the two objects have such similar brightnesses. Sila, the "primary", was not always brighter, but averaged over the eight visits when separate photometry was obtained in filters approximating *V* band, it was brighter by a mean of 0.12 mags.

Figure 2 compares observed astrometry and positions from our best-fit Keplerian orbital solution for Nunam's motion relative to Sila. The fitted elements of this solution are listed in Table 2 along with their 1-σ uncertainties. These uncertainties were estimated by randomly generating 1000 altered sets of astrometric observations by adding Gaussian noise to the real observations consistent with their error bars. A new orbit was fitted to each of these altered data sets, leading to a collection of Monte Carlo orbits that we used to characterize the probability distribution for each fitted element. Table 2 also lists derived quantities like the system mass $M_{sys}$, and Hill radius $r_H$, computed from the mutual orbit elements.

Table 2 shows an eccentricity that is non-zero by more than its 1-σ uncertainty. To further explore the possibility of a circular orbit we performed a restricted fit in which the eccentricity was forced to be exactly zero, resulting in an orbit solution with $\chi^2 = 33.8$. This solution can only be excluded at 1.8-σ confidence, meaning that it cannot yet be ruled out, especially considering that noise in observations of a circular orbit tends to produce non-zero apparent eccentricities in fitting the data (e.g., Buie et al. 2012). This seems especially risky where data from diverse instruments are combined, as done here. Potential errors in the plate scale of an instrument could manifest themselves as a bogus eccentricity or otherwise distort the orbit solution. For STIS, NICMOS/NIC2, ACS/HRC, WFPC2/PC, and WFC3/UVIS, plate scales are reported to 4, 4, 3, 4, and 4 decimals, respectively (e.g., http://www.stsci.edu/hst/HST_overview/documents/multidrizzle/ch43.html).

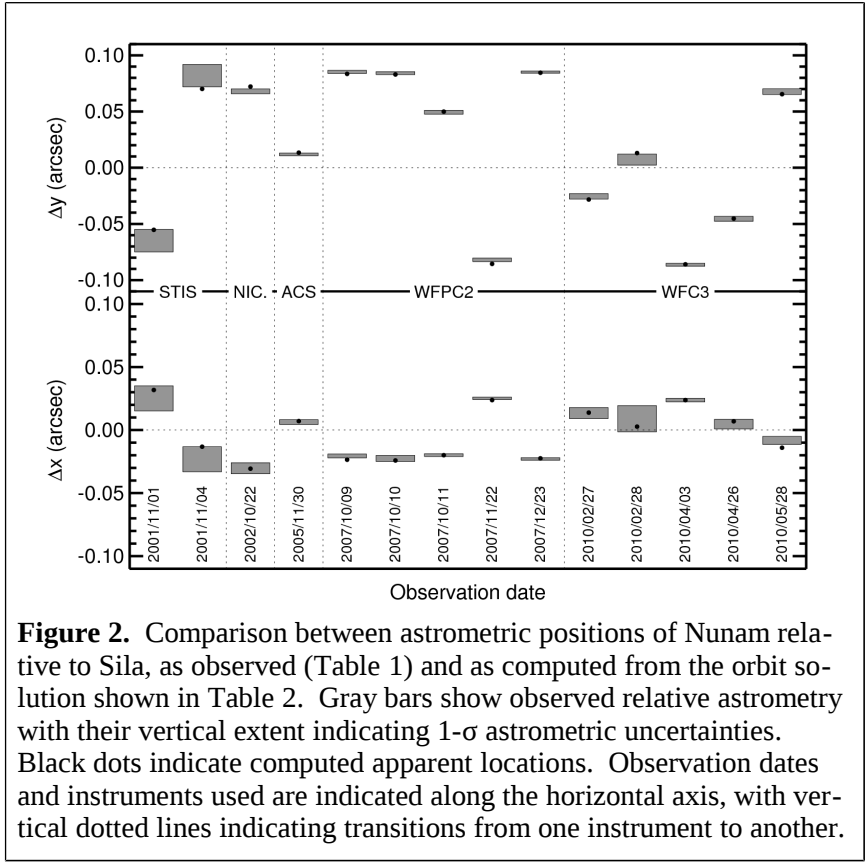

**Figure 2.** Comparison between astrometric positions of Nunam relative to Sila, as observed (Table 1) and as computed from the orbit solution shown in Table 2. Gray bars show observed relative astrometry with their vertical extent indicating 1-σ astrometric uncertainties. Black dots indicate computed apparent locations. Observation dates and instruments used are indicated along the horizontal axis, with vertical dotted lines indicating transitions from one instrument to another.

Since the fractional uncertainties on the reported astrometric positions in Table 1 are always worse than 1%, it seems unlikely that plate scale uncertainties could contribute a significant additional source of error.

**Table 2**
Mutual orbital elements, derived parameters, and 1-σ uncertainties.

| Parameter | | Value |
|---|---|---|
| Fitted elements of mutual orbit:[a] | | |
|    Period (days) | $P$ | 12.50995 ± 0.00036 |
|    Semimajor axis (km) | $a$ | 2777 ± 19 |
|    Eccentricity | $e$ | 0.020 ± 0.015 |
|    Inclination[b] (deg) | $i$ | 103.51 ± 0.39 |
|    Mean longitude[b] at epoch[c] (deg) | $\varepsilon$ | 16.3 ± 1.0 |
|    Longitude of asc. node[b] (deg) | $\Omega$ | 140.76 ± 0.66 |
|    Longitude of periapsis[b] (deg) | $\varpi$ | 326 ± 59 |
| Derived parameters: | | |
|    Standard gravitational parameter $GM_{sys}$ (km$^3$ day$^{-2}$) | $\mu$ | 0.724 ± 0.015 |
|    System mass[d] (10$^{18}$ kg) | $M_{sys}$ | 10.84 ± 0.22 |
|    Hill radius[d] (10$^3$ km) | $r_H$ | 766.5 ± 5.3 |
| Orbit pole right ascension[b] (deg) | $\alpha_{pole}$ | 50.76 ± 0.67 |
| Orbit pole declination[b] (deg) | $\delta_{pole}$ | –13.51 ± 0.39 |
| Orbit pole ecliptic longitude (deg) | $\lambda_{pole}$ | 44.19 ± 0.78 |
| Orbit pole ecliptic latitude (deg) | $\beta_{pole}$ | –30.92 ± 0.34 |
| Inclination between mutual orbit and mean heliocentric orbit (deg) | | 120.05 ± 0.35 |

Table notes:

[a.] Elements are for orbital motion of Nunam relative to Sila. This solution has $\chi^2 = 31.8$, corresponding to reduced $\chi_v^2 = 1.51$ if astrometric observations from all 14 epochs are independent of one another, our estimated astrometric uncertainties are correct, and they obey a Gaussian distribution. The mirror orbit has $\chi^2 = 38.6$, and is excluded at 99% confidence, subject to these assumptions. The observations are consistent with zero eccentricity, so the longitude of periapsis $\varpi$ is poorly constrained.

[b.] Referenced to J2000 equatorial frame.

[c.] The epoch is Julian date 2454400.0 (2007 October 26 12:00 UT).

[d.] Computed as $M_{sys} = \frac{4\pi^2 a^3}{P^2 G}$ and $r_H = a_\odot (1-e_\odot)(\frac{M_{sys}}{3M_\odot})^{1/3}$ (Hamilton and Burns 1992), where $G$ is the gravitational constant ($6.67428 \times 10^{-8}$ g$^{-1}$ s$^{-2}$ cm$^3$, Mohr et al. 2008), $a_\odot$ and $e_\odot$ are the mean semi-major axis and eccentricity of the heliocentric orbit, and $M_\odot$ is the mass of the Sun.

## 4. Mutual Events

A particularly noteworthy feature of the mutual orbit of Sila and Nunam is that during the present epoch, observers located in the inner solar system are viewing it nearly perfectly edge-on. This orientation means that Sila and Nunam alternate in passing in front of one another as seen from Earth, much like the mutual events of Pluto and Charon during the 1980s that enabled observers to make tremendous gains in knowledge about that system (e.g., Binzel and Hubbard 1997). To review mutual event nomenclature, "superior" events are when the primary body is in the foreground and "inferior" events are when the secondary is in the foreground. Occultation type events occur when the foreground object obstructs the view of part of the other body. In eclipse type events, the shadow cast by the foreground object impinges on the other body. Since the Sun and Earth are always separated by small angles as seen from Sila and Nunam's distant location at ~44 AU, most events combine both eclipses and occultations. The plane of Sila and Nunam's mutual orbit is highly inclined to that of its heliocentric orbit (120 degrees), so mutual events only occur during brief seasons twice each 3-century heliocentric orbit, when the heliocentric orbit sweeps the mutual orbit plane across the inner solar system. In addition to the inclination between orbit planes, the duration of these mutual event seasons depends on the sizes of the bodies: the larger they are relative to their separation, the longer the seasons last. For nominal sizes from thermal observations (see Section 6), the event season should last about a decade.

The seasonal evolution of eclipse geometry is relatively simple, since it depends only on the binary's mutual and heliocentric orbits. Early in the season, the northern hemisphere of the background body gets shadowed (using the spin vector to define North, and assuming the bodies are tidally synchronized with their spin axes coinciding with the orbit pole, as we will argue in Section 5). Later, events become more central, and then finally, only the southern hemisphere of the background body is shadowed. This trend is illustrated in Fig. 3, where the shadow (indicated by cross-hatching) is seen to gradually shift from West to East on the sky plane as the season unfolds. The seasonal evolution of occultation geometry is somewhat more complex, since it also depends on the location of the observer. For an Earth-based observer, at western quadrature (early in each annual apparition, top row in Fig. 3) Sila and Nunam are 90 degrees West of the Sun in the sky, and the foreground body's shadow extends to the West. As Earth moves around the Sun, the angle between Earth and Sun as seen from Sila and Nunam closes, until the shadow is mostly hidden behind the foreground object at opposition (middle row in Fig. 3). At eastern quadrature, late in the apparition, the foreground object's shadow extends to the East (bottom row in Fig. 3).

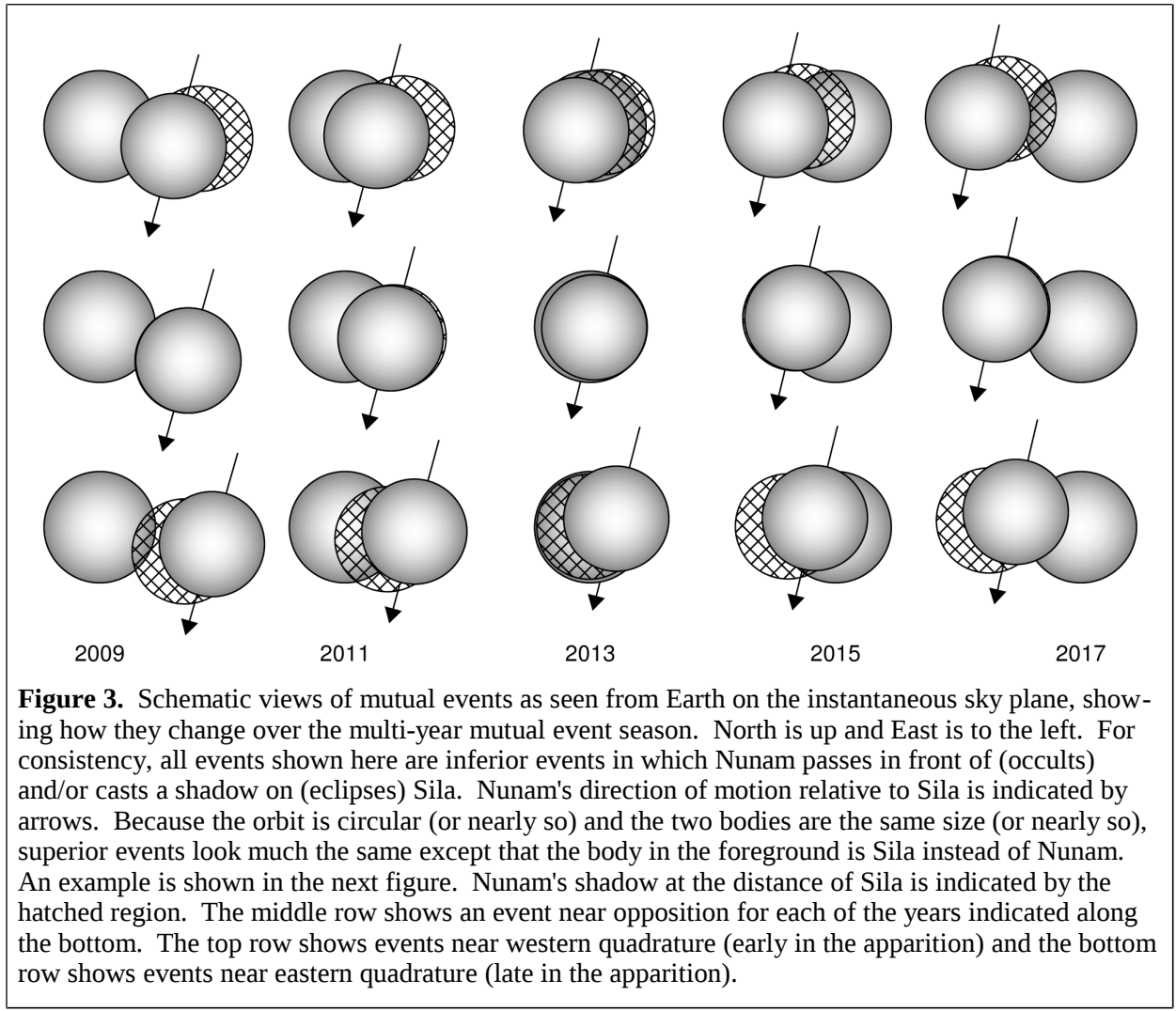

**Figure 3.** Schematic views of mutual events as seen from Earth on the instantaneous sky plane, showing how they change over the multi-year mutual event season. North is up and East is to the left. For consistency, all events shown here are inferior events in which Nunam passes in front of (occults) and/or casts a shadow on (eclipses) Sila. Nunam's direction of motion relative to Sila is indicated by arrows. Because the orbit is circular (or nearly so) and the two bodies are the same size (or nearly so), superior events look much the same except that the body in the foreground is Sila instead of Nunam. An example is shown in the next figure. Nunam's shadow at the distance of Sila is indicated by the hatched region. The middle row shows an event near opposition for each of the years indicated along the bottom. The top row shows events near western quadrature (early in the apparition) and the bottom row shows events near eastern quadrature (late in the apparition).

For the orbit solution in Table 2, the deepest, most central events will be observable during the 2013 apparition, for which detailed timings are tabulated in Table 3. However, the orbital parameters have associated uncertainties. The effect of these uncertainties on event geometry is illustrated in Fig. 4, where we have used the Monte Carlo cloud of 1000 orbits fitted to randomized versions of the astrometric observations described earlier to illustrate the probability that Nunam would be in a somewhat different place relative to Sila during the nominal mid-time of the 2011 February 1 UT event. Without additional observations, this cloud would gradually expand over time, mostly parallel to the direction of relative motion indicated by the arrow. Event observations with good time resolution can collapse the orbital uncertainties that contribute to the extent of this probability cloud.

During the 2011 apparition, we attempted to observe three events at Lowell Observatory

(on Feb. 1, Feb. 26, and Mar. 23 UT) but only one of these events produced useable data, due to poor weather. One of us (L. Wasserman) obtained a dozen consecutive 900 second integrations through partly-cloudy sky conditions at Lowell Observatory's 1.8 m Perkins telescope at Anderson Mesa on Feb. 1 UT. These observations made use of the Perkins Re-Imaging SysteM (PRISM; Janes et al. 2004) equipped with a 2048 × 2048 pixel Fairchild CCD and a broadband $V+R$ filter (0.5 to 0.7 μm). The variable sky conditions prevented us from doing absolute photometry, but photometry relative to an average of several nearby field stars revealed an apparent dip of approximately 0.2 mag during the portion of the event observed.

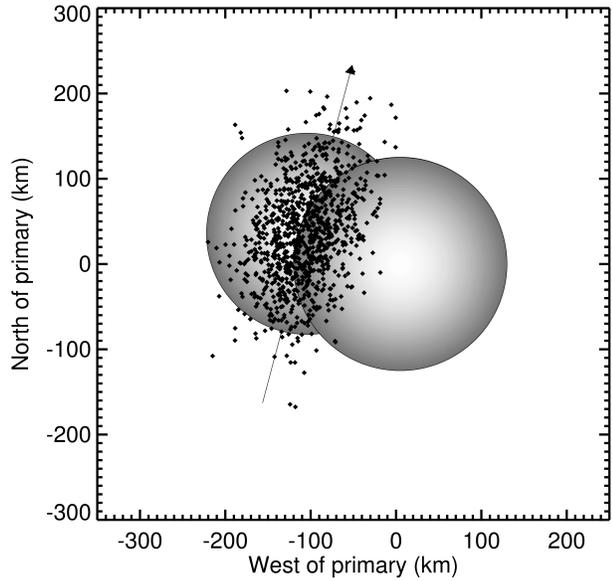

**Figure 4.** Effect of uncertainties in the mutual orbit illustrated on a schematic drawing of the 2011 February 1 UT superior event. Small black spots are relative locations of Nunam computed for orbits fitted to 1000 randomized sets of astrometric data. When projected to the sky plane at the time of an event, the effect of uncertainty in the mutual orbit produces smear, mostly along the direction of relative motion, translating to uncertainty in event timing. The smaller projected uncertainty in the cross-track direction translates to uncertainty in the timing of the beginning, middle, and end of the mutual event season, and to the depths and durations of individual events. Sila's shadow is not seen because this particular event coincides with opposition.

**Table 3**
Predicted UT dates of mutual events during the 2012-2013 apparition.

| First contact | Minimum light | Last contact | Event type |
|---|---|---|---|
| 2012/10/24 15:41 | 2012/10/24 20:04 | 2012/10/25 00:22 | Inf. |
| 2012/10/30 22:21 | 2012/10/31 02:55 | 2012/10/31 07:25 | Sup. |
| 2012/11/06 03:52 | 2012/11/06 08:15 | 2012/11/06 12:33 | Inf. |
| 2012/11/12 10:32 | 2012/11/12 15:09 | 2012/11/12 19:37 | Sup. |
| 2012/11/18 16:04 | 2012/11/18 20:27 | 2012/11/19 00:45 | Inf. |
| 2012/11/24 22:45 | 2012/11/25 03:22 | 2012/11/25 07:48 | Sup. |
| 2012/12/01 04:18 | 2012/12/01 08:40 | 2012/12/01 12:56 | Inf. |
| 2012/12/07 11:00 | 2012/12/07 15:31 | 2012/12/07 20:00 | Sup. |
| 2012/12/13 16:31 | 2012/12/13 20:50 | 2012/12/14 01:08 | Inf. |
| 2012/12/19 23:15 | 2012/12/20 03:44 | 2012/12/20 08:11 | Sup. |
| 2012/12/26 04:48 | 2012/12/26 09:07 | 2012/12/26 13:20 | Inf. |
| 2013/01/01 11:33 | 2013/01/01 15:56 | 2013/01/01 20:24 | Sup. |
| 2013/01/07 17:05 | 2013/01/07 21:20 | 2013/01/08 01:32 | Inf. |
| 2013/01/13 23:51 | 2013/01/14 04:09 | 2013/01/14 08:36 | Sup. |
| 2013/01/20 05:25 | 2013/01/20 09:33 | 2013/01/20 13:45 | Inf. |
| 2013/01/26 12:11 | 2013/01/26 16:28 | 2013/01/26 20:48 | Sup. |
| 2013/02/01 17:45 | 2013/02/01 21:54 | 2013/02/02 01:59 | Inf. |
| 2013/02/08 00:30 | 2013/02/08 04:43 | 2013/02/08 09:03 | Sup. |
| 2013/02/14 06:00 | 2013/02/14 10:07 | 2013/02/14 14:17 | Inf. |
| 2013/02/20 12:43 | 2013/02/20 16:56 | 2013/02/20 21:24 | Sup. |
| 2013/02/26 18:12 | 2013/02/26 22:21 | 2013/02/27 02:37 | Inf. |
| 2013/03/05 00:56 | 2013/03/05 05:13 | 2013/03/05 09:44 | Sup. |
| 2013/03/11 06:26 | 2013/03/11 10:36 | 2013/03/11 14:57 | Inf. |
| 2013/03/17 13:10 | 2013/03/17 17:26 | 2013/03/17 22:03 | Sup. |
| 2013/03/23 18:42 | 2013/03/23 22:54 | 2013/03/24 03:14 | Inf. |
| 2013/03/30 01:26 | 2013/03/30 05:41 | 2013/03/30 10:21 | Sup. |
| 2013/04/05 06:56 | 2013/04/05 11:10 | 2013/04/05 15:33 | Inf. |
| 2013/04/11 13:40 | 2013/04/11 17:59 | 2013/04/11 22:39 | Sup. |
| 2013/04/17 19:10 | 2013/04/17 23:25 | 2013/04/18 03:49 | Inf. |
| 2013/04/24 01:54 | 2013/04/24 06:13 | 2013/04/24 10:55 | Sup. |
| 2013/04/30 07:26 | 2013/04/30 11:39 | 2013/04/30 16:06 | Inf. |
| 2013/05/06 14:10 | 2013/05/06 18:29 | 2013/05/06 23:11 | Sup. |

As part of a separate study of photometric phase behavior (see next section), another of us (D. Rabinowitz), observed the system in queue mode at the Small and Moderate Aperture Research Telescope System (SMARTS) 1.3 m telescope at Cerro Tololo, Chile, on the same night. These observations consisted of six non-consecutive 600 second integrations through a Johnson $R$ filter (0.59 to 0.72 μm) using A Novel Double-Imaging CAMera (ANDICAM), also equipped with a 2048 × 2048 pixel Fairchild CCD. Details of data analysis procedures were described by Rabinowitz et al. (2007). The SMARTS data show flux changes very similar to those shown by the Perkins observations.

Both sets of observations are tabulated in Table 4. The data are consistent with the timing and the magnitude of the flux dip predicted by assuming Sila and Nunam are spheres with Lambertian scattering behavior and radii of 125 and 118 km, as shown in Fig. 5. The observations seem to confirm that the predicted mutual event season is underway, although the combined set of observations has insufficient signal precision and temporal resolution to appreciably tighten constraints on the mutual orbit or the sizes of the individual bodies, especially considering that we do not yet know their spin states or lightcurves.

Anticipating more and better mutual event data in the future, we ran models in which we varied the radii of Sila and Nunam by ±5 km, to see what sort of effect that had on model event lightcurves. All other things being equal, changing the size of either object resulted in changes in observable flux during an event by up to several percent. The temporal pattern of these flux differences was quite distinctive, depending which object's size was changed, whether the event modeled was a superior or an inferior event, and on the relative importance of occultation and eclipse components of the event (these vary over the course of an apparition as shown in Fig. 3). These preliminary models lead us to expect that observations with achievable temporal resolution ($\Delta t <$ 10 min) and signal precision (S/N > 50) could determine the individual radii of Sila and Nunam to precisions of a few km. However, these sizes would necessarily be subject to assumptions regarding the spin states of the objects, the circularity of their limb profiles, and their center-to-limb photometric brightness profiles, as well as requiring greatly improved knowledge of the mutual orbit. Multiple event observations would be required to simultaneously constrain the sizes along with these other parameters.

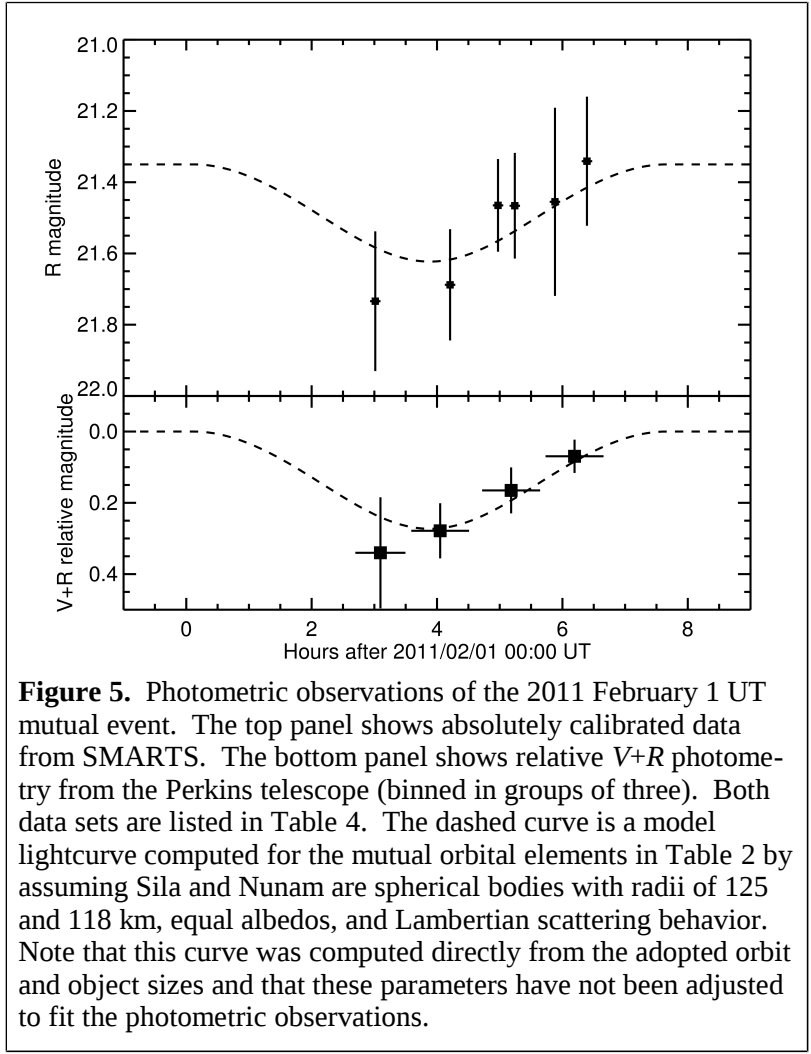

**Figure 5.** Photometric observations of the 2011 February 1 UT mutual event. The top panel shows absolutely calibrated data from SMARTS. The bottom panel shows relative $V+R$ photometry from the Perkins telescope (binned in groups of three). Both data sets are listed in Table 4. The dashed curve is a model lightcurve computed for the mutual orbital elements in Table 2 by assuming Sila and Nunam are spherical bodies with radii of 125 and 118 km, equal albedos, and Lambertian scattering behavior. Note that this curve was computed directly from the adopted orbit and object sizes and that these parameters have not been adjusted to fit the photometric observations.

**Table 4**
Observations of the 2011-02-01 UT mutual event.

| UT Mid-time | Telescope | Filter | Photometry (mags) |
| --- | --- | --- | --- |
| 2:50 | Perkins | $V+R$ | 0.42 ± 0.22 |
| 3:05 | Perkins | $V+R$ | 0.14 ± 0.23 |
| 3:23 | Perkins | $V+R$ | 0.29 ± 0.22 |
| 3:43 | Perkins | $V+R$ | 0.22 ± 0.19 |
| 4:03 | Perkins | $V+R$ | 0.28 ± 0.14 |
| 4:23 | Perkins | $V+R$ | 0.23 ± 0.09 |
| 4:50 | Perkins | $V+R$ | 0.05 ± 0.08 |
| 5:11 | Perkins | V+R | 0.14 ± 0.12 |
| 5:31 | Perkins | V+R | 0.27 ± 0.12 |
| 5:51 | Perkins | V+R | 0.08 ± 0.10 |
| 6:12 | Perkins | V+R | 0.04 ± 0.07 |
| 6:32 | Perkins | V+R | 0.00 ± 0.07 |
| 3:01 | SMARTS | R | 21.73 ± 0.20 |
| 4:13 | SMARTS | R | 21.69 ± 0.16 |
| 4:58 | SMARTS | R | 21.47 ± 0.13 |
| 5:14 | SMARTS | R | 21.47 ± 0.15 |
| 5:53 | SMARTS | R | 21.46 ± 0.26 |
| 6:23 | SMARTS | R | 21.34 ± 0.18 |

Table note: The SMARTS *R* observations are absolutely calibrated. The Perkins *V+R* observations are not, with photometry relative to field stars having been arbitrarily scaled so the last and brightest observation has a magnitude of zero. At the time of these observations, the distance from Sun to Sila and Nunam was $r = 43.502$ AU, and from Earth to Sila and Nunam was $\Delta = 42.516$ AU. The phase angle *g* ranged from 0.007° to 0.008°.

## 5. Spin State and Ground-Based Photometry

A tight, circular orbit like we have found for Sila and Nunam is the optimal configuration for a doubly-synchronous state; i.e. the rotational periods of the two objects being exactly equal to their mutual orbital period and their spin poles coinciding with the orbit pole. With a mean separation of 2780 km (corresponding to 0.0036 Hill radii, or about 22 Sila radii), Sila and Nunam are tightly bound in a manner comparable to Pluto and Charon (at 0.0032 Hill radii, or 17 Pluto radii). Since Pluto and Charon are doubly synchronous (as confirmed by Buie et al. 1997), it seems likely that Sila and Nunam could be as well. To be more quantitative, we calculated the approximate timescale to reach that state from a much faster initial rotation rate. For an approximately spherical object in a circular orbit, the timescale to spin down to synchronous rotation is (Gladman et al. 1996):

$$\tau_{despin} = \frac{16\pi}{15} \frac{a^6 \rho}{G M_o^2 T_i} \frac{Q}{k_2}, \tag{1}$$

where *a* is the semimajor axis of the mutual orbit, $\rho$ is the density of the object in question, $M_o$ is the mass of the other object, and $T_i$ is the initial rotational period. The $Q/k_2$ term defines the dimensionless speed of tidal dissipation in the object. Assuming bulk densities of 0.72 g cm$^{-3}$ (see next section), near equal masses, and a tidal $Q = 100$, similar to the irregular satellites of Neptune (Zhang and Hamilton 2008), the slowest tidal evolution would be for a rigid, icy body, which would have a $Q/k_2 \approx 10^6$ (Burns 1977; Gladman et al. 1996). For Sila and Nunam's mutual orbit, this gives the relation $\tau_{despin} T_i = 6.7 \times 10^{20}$ sec$^2$ (or, in more convenient units, $5.9 \times 10^9$ yr hr). Typical solitary TNO rotation periods range from 4 to 24 hours (Sheppard et al. 2008; Thirouin et al. 2010), giving a range for $\tau_{despin}$ of 0.2-1.5 Ga. Thus, even if the objects had relatively fast rotation rates immediately after formation ~4 Ga ago, they should be completely despun to doubly synchronous rotation by the present time.

In addition to principal-axis rotation, the objects could also have rotations excited by impacts not along their principal axes. The timescale to damp out this perturbed motion is (Gladman et al. 1996):

$$\tau_{\text{wobble}} = \frac{G \rho T_i^3}{5\pi} \frac{Q}{k_2}. \tag{2}$$

If the system is already synchronized so $T_i$ is the orbital period, this timescale is of the order of 0.1 Ga. Therefore, after Sila and Nunam have arrived in the doubly synchronous state, they are relatively robust to rotational perturbations caused by impacts, and we would expect them to be fully tidally locked today.

To check for possible long-period photometric variations as expected for a tidally locked system, we obtained additional ground-based photometry at four different telescopes. Observations by D. Rabinowitz during 16 additional nights made use of ANDICAM and an *R* filter on the SMARTS 1.3 m telescope, as described in the previous section. S. Benecchi observed the system on 5 additional nights using the Carnegie Institution for Science's 2.5 m Irénée du Pont telescope at Las Campanas. These observations made use of a SITe2k CCD camera and a Sloan *r'* filter (0.56-0.69 µm). The du Pont data were normalized to the SMARTS data. A. Verbiscer observed the system on 27 January 2011 UT using the Seaver Prototype Imaging camera (SPIcam) on the 3.5 m Astrophysical Research Consortium (ARC) telescope at Apache Point Observatory (APO). SPIcam is equipped with a backside-illuminated SITe TK 2048E 2048 × 2048 pixel CCD. Two 300 second integrations obtained through a broad *R* filter (0.50-0.80 µm) were recorded. Additionally, Verbiscer observed the system on 3 nights using the VATT4k Imager at the 1.8m Vatican Advanced Technology Telescope (VATT). The VATT4k is a STA0500A 4064 × 4064 pixel CCD. Multiple, consecutive 600-second integrations were obtained through a Harris *R* filter (0.51-0.75 µm) on each night. Nightly average *R*-band photometric measurements from these observations appear in Table 5, reduced to absolute magnitude $H_R$ to remove effects of changing Sun-object and observer-object distances (*r* and *Δ*). We saw no convincing evidence in the SMARTS data for a solar phase effect, so no correction was applied for phase angle *g*. These data are too heterogeneous and of insufficient quality to uniquely derive a period, but if the system were tidally locked with the long axes of both bodies pointed toward one another as expected, we would see a double-peaked lightcurve with a period matching the orbital period of 12.50995 ± 0.00036 days, or equivalently, a single-peaked lightcurve with half this period. Phasing the data to 6.254975 days produces what looks like the expected single-peaked lightcurve, as shown in Fig. 6. A sinusoid fitted to the phased SMARTS photometry has a peak-to-peak amplitude of 0.14 ± 0.07 mags oscillating about a mean reduced *R* magnitude $H_R$ = 4.93 ± 0.02, with a statistically insignificant phase shift of 0.03 ± 0.07. An 0.14 mag lightcurve corresponds to an axis ratio of *c*/*a* = 1.14, if both bodies share the same prolate ellipsoidal figure where *c* is the long axis and *a* = *b* are the short axes. If only one of the two bodies is responsible for the apparent lightcurve variation, its axis ratio would have to be about 1.28 to produce the same total photometric variation for the system.

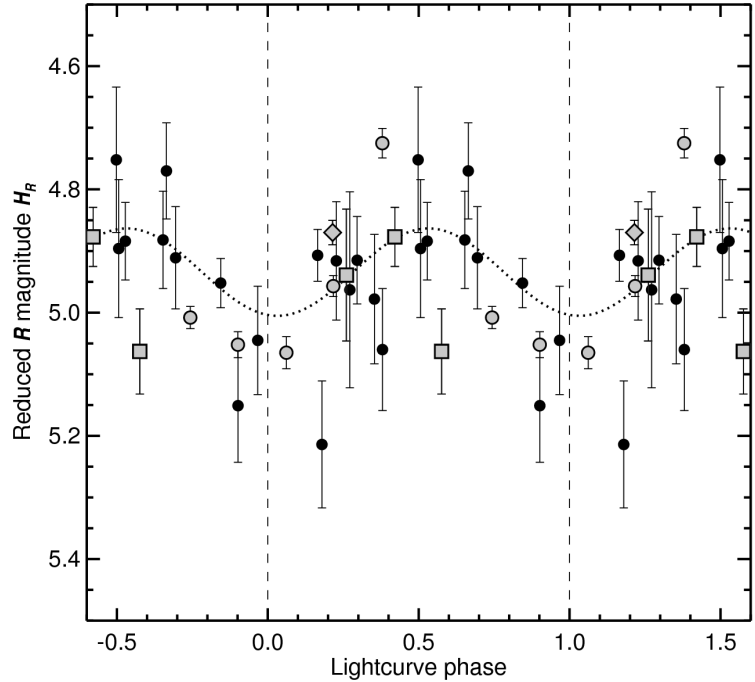

**Figure 6.** Phased lightcurve for the photometry in Table 5, assuming a period of 6.254975 days, exactly half of the orbital period. If, the system is tidally locked as expected, this phasing would give a single-peaked lightcurve over the phase interval 0 to 1. Zero phase is defined as the phase where mutual events occur. Data are duplicated outside the 0 to 1 interval to better show the periodic variation. The dotted curve is a sine function fitted to just the SMARTS data (solid black points, the largest internally consistent data set), resulting in a mean magnitude of 4.93 ± 0.02 and a peak-to-peak amplitude of 0.14 ± 0.07 mags. Points with gray centers, not used in the sinusoid fit, are from du Pont (gray circles, scaled to the SMARTS data), ARC (gray diamond), and VATT (gray squares).

**Table 5**
Non-event nightly-average *R*-band photometry of combined flux from Sila and Nunam.

| Mean UT time | Telescope | *r* (AU) | *Δ* (AU) | Phase angle (deg.) | Reduced *R* magnitude |
|---|---|---|---|---|---|
| 2010/12/15 8:00 | SMARTS | 43.502 | 42.855 | 0.98 | 4.98 ± 0.10 |
| 2010/12/17 6:41 | SMARTS | 43.502 | 42.830 | 0.95 | 4.77 ± 0.08 |
| 2010/12/27 8:00 | SMARTS | 43.502 | 42.713 | 0.78 | 4.96 ± 0.16 |
| 2010/12/31 6:31 | SMARTS | 43.502 | 42.673 | 0.70 | 5.15 ± 0.09 |
| 2011/01/02 7:25 | SMARTS | 43.502 | 42.655 | 0.66 | 4.92 ± 0.10 |
| 2011/01/03 6:23 | SMARTS | 43.502 | 42.646 | 0.65 | 5.06 ± 0.10 |
| 2011/01/05 5:35 | SMARTS | 43.502 | 42.630 | 0.60 | 4.91 ± 0.08 |
| 2011/01/08 6:23 | SMARTS | 43.502 | 42.606 | 0.54 | 5.21 ± 0.10 |
| 2011/01/10 7:25 | SMARTS | 43.502 | 42.592 | 0.50 | 4.90 ± 0.11 |
| 2011/01/15 6:01 | SMARTS | 43.502 | 42.562 | 0.39 | 4.91 ± 0.07 |
| 2011/01/29 5:05 | SMARTS | 43.502 | 42.518 | 0.07 | 4.88 ± 0.06 |
| 2011/01/31 4:30 | SMARTS | 43.502 | 42.516 | 0.02 | 4.95 ± 0.04 |
| 2011/02/02 4:43 | SMARTS | 43.502 | 42.516 | 0.03 | 4.91 ± 0.04 |
| 2011/02/04 6:41 | SMARTS | 43.501 | 42.517 | 0.10 | 4.75 ± 0.12 |
| 2011/02/05 5:56 | SMARTS | 43.501 | 42.518 | 0.10 | 4.88 ± 0.08 |
| 2011/02/07 5:03 | SMARTS | 43.501 | 42.521 | 0.14 | 5.05 ± 0.09 |
| 2011/01/27 6:03 | ARC | 43.502 | 42.521 | 0.11 | 4.87 ± 0.02 |
| 2011/03/09 2:06 | du Pont | 43.501 | 42.704 | 0.79 | 5.01 ± 0.02 |
| 2011/03/10 1:48 | du Pont | 43.501 | 42.714 | 0.81 | 5.05 ± 0.02 |
| 2011/03/11 1:43 | du Pont | 43.501 | 42.725 | 0.82 | 5.07 ± 0.03 |
| 2011/03/12 1:15 | du Pont | 43.501 | 42.735 | 0.84 | 4.96 ± 0.02 |
| 2011/03/13 1:39 | du Pont | 43.501 | 42.747 | 0.86 | 4.72 ± 0.02 |
| 2011/10/23 12:03 | VATT | 43.498 | 43.733 | 1.27 | 4.94 ± 0.11 |
| 2011/10/24 12:09 | VATT | 43.498 | 43.716 | 1.28 | 4.88 ± 0.05 |
| 2011/10/25 11:23 | VATT | 43.498 | 43.699 | 1.28 | 5.06 ± 0.07 |

Table note: Photometric observations reduced to absolute magnitude to remove effects of varying distances between the Sun and object ($r$) and between the observer and object ($\Delta$). The du Pont data, taken with a Sloan $r'$ filter, were scaled to match the SMARTS data.

## 6. Size and Density

Vilenius et al. (2012) detected thermal emission from Sila and Nunam using the Photodetector Array Camera and Spectrometer (PACS; Poglitsch et al. 2010) of the 3.5 m Herschel Space Observatory, located at the Earth's L2 Lagrange point. They combined their observations with earlier Spitzer and Herschel observations (Stansberry et al. 2008; Müller et al. 2010), to estimate that the projected surface area of the system corresponds to that of a single spherical body with effective diameter $D_{\text{eff}} = 343^{+41}_{-42}$ km. After checking that none of the thermal observations coincided with mutual event times, we combined the Vilenius et al. size with our visual wavelength photometry to conclude that the average geometric albedo of the system $A_p$ is $0.086^{+0.026}_{-0.017}$ at $V$ band (based on HST data in Table 1) and $0.117^{+0.035}_{-0.024}$ at $R$ band (based on SMARTS data in Table 5) assuming –26.74 and –27.10 mags for solar brightness at 1 AU at $V$ and $R$, respectively. Assuming Sila and Nunam both have the same albedo (consistent with the similar colors among most other binary pairs reported by Benecchi et al. 2009), our observed mean magnitude difference between them ($\Delta_{\text{mag}} = 0.12$) translates into mean radii of the two bodies of 125 ± 15 and 118 ± 14 km, respectively. Using these sizes, along with our system mass of (10.84 ± 0.22) × $10^{18}$ kg, and assuming spherical shapes and equal densities for the two bodies, we can compute a bulk density of $0.72^{+0.37}_{-0.22}$ g cm$^{-3}$. If instead of spherical shapes, we assume Sila and Nunam are prolate ellipsoids with axial ratio $c/a = 1.14$ and the same surface area as the equivalent spheres averaged over their rotations, Sila would have $c = 137^{+16}_{-17}$ km and $a = 120^{+14}_{-15}$ km, and Nunam would have $c = 130 \pm 16$ km and $a = 114 \pm 14$ km, resulting in a total volume about 3% larger and a comparably reduced bulk density of $0.70^{+0.36}_{-0.21}$ g cm$^{-3}$. To account for this possibility, we extend the lower error bar of our earlier estimate and adopt density $0.72^{+0.37}_{-0.23}$ g cm$^{-3}$.

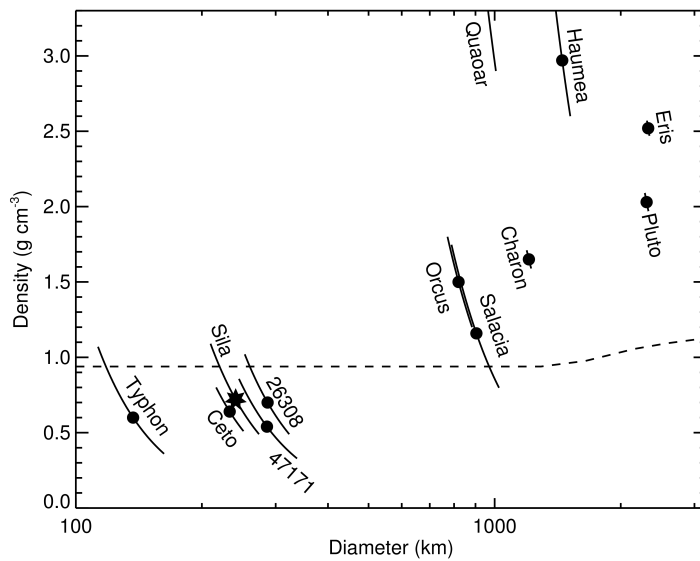

**Figure 7.** Estimated bulk density and diameter of Sila (indicated by a star) compared with values for other Kuiper belt objects with accurate masses from satellite orbits (circles; from Buie et al. 2006; Rabinowitz et al. 2006; Brown and Schaller 2007; Benecchi et al. 2010; Fraser and Brown 2010; Sicardy et al. 2011; Santos-Sanz et al. 2012; Stansberry et al. 2012; and Vilenius et al. 2012). The dashed curve represents the density of a pure ice sphere showing the effect of gravitational self compression as a function of size (Lupo and Lewis 1979). The low densities of smaller objects implies the presence of appreciable void space within them, while the higher densities of larger objects requires denser materials such a silicates.

This density is comparable to other densities of small binary TNBs, as shown in Fig. 7, despite the other small objects belonging to more excited dynamical classes. The larger bodies, which also belong to more excited dynamical classes, have generally higher bulk densities. The density contrast between smaller and larger TNBs is greater than can be explained by gravitational self-compression, arguing for distinct compositions. Whether these compositional differences result from the larger bodies having accreted in compositionally distinct nebular environments, or from post-differentiation impact stripping of lower density mantle materials (e.g., Brown 2010) remains unknown. This small sample of known densities includes a wide range of dynamical classes, and density uncertainties remain substantial for many of them. More and better data will be needed to disentangle possible effects of size, nebular source, and subsequent evolutionary history. At least for Sila and Nunam, we can anticipate better size constraints from mutual event observations in the near future, leading to tighter constraints on albedo and bulk density.

## 7. Summary and Discussion

We report results of an observational study of a tight transneptunian binary system with near-equal brightness components. Sila and Nunam were spatially resolved in Hubble Space Telescope images taken with five different instruments from 2001 through 2010. Relative astrometry from the Hubble observations enabled us to determine the circular or near-circular, retrograde, orbit of Nunam relative to Sila with a period of 12.50995 ± 0.00036 days and a semimajor axis of 2777 ± 19 km. The orbit is fortuitously oriented such that its plane is currently sweeping across the inner solar system, causing the system to undergo mutual events in which Sila and Nunam alternate in passing between one another and the Sun and Earth. We observed part of one of these events from two different moderate-aperture, ground-based telescopes, seeing a dip in flux at the expected time in both sets of observations. Future mutual event observations with improved signal precision and temporal resolution have the potential to reveal a great deal of information about the system. The timing and depth of mutual events can be used to better constrain the orbital parameters and the sizes of the two bodies. Observations of numerous events can be used to detect albedo features on the faces of the two bodies oriented toward one another (assuming they are indeed tidally locked) and also to probe whether or not the limb profiles as seen from Earth at event times are circular, as we have assumed they are ($b = c$). Observations when one of the two objects is hidden from view by the other can be used to separately study the individual properties of the two bodies using instrumentation that cannot spatially resolve them, such as spectrometers and thermal infrared radiometers. According to our current best orbit solution, the optimal time for such studies is the 2013 apparition, although additional data are needed to refine the orbital parameters to verify that this is indeed the best observing season for this purpose.

Knowledge of the mutual orbit of Sila and Nunam also enables calculation of the total mass of the system, $(10.84 \pm 0.22) \times 10^{18}$ kg. Combining this mass with estimates of the projected area of the system from thermal infrared observations by Spitzer and Herschel Space Observatories enables estimation of the average bulk density of the system as $0.72^{+0.37}_{-0.23}$ g cm$^{-3}$. This is the first bulk density for a Cold Classical TNO based on an accurate system mass from a satellite orbit. Cold Classical TNOs are of particular interest in that they are thought to have formed in the outermost part of the protoplanetary disk, beyond the giant planet forming region where the more dynamically excited TNO populations are thought to have originally formed. Having formed and remained so far from the Sun, Cold Classicals are thought to preserve solids from that environment in a relatively unaltered state. The bulk density computed for Sila and Nunam is comparable to bulk densities of similarly sized binary TNOs from the more dynamically excited populations, but the low bulk densities of small TNOs contrast with the much larger densities of planet-sized TNOs, implying distinct compositions with higher rock fractions in the larger bodies. Of course, an important caveat is that the sample of known TNO bulk densities remains very small, and includes objects from a broad mix of dynamical classes, that, as binaries, may not even be particularly representative of the single objects on similar heliocentric orbits. Additionally, as one of the brightest probable members of the Cold Classical population, Sila and Nunam may not be particularly representative of the numerous fainter Cold Classical TNOs.

## Acknowledgments

This work is based in part on NASA/ESA Hubble Space Telescope programs 9110, 9386, 10514, 11178, and 11650. The Space Telescope Science Institute (STScI) is operated by the Association of Universities for Research in Astronomy, Inc., under NASA contract NAS 5-26555. Support for programs 11178 and 11650 was provided by NASA through grants from STScI. We are especially grateful to Tony Roman at STScI for his quick action in scheduling HST follow-up observations. This work was also based in part on data from the SMARTS telescopes, operated by the SMARTS consortium, from the VATT: the Alice P. Lennon Telescope and the Thomas J. Bannan Astrophysics Facility, from Lowell Observatory's Perkins telescope operated as a partnership between Lowell Observatory, Boston University, and Georgia State University, and from the Apache Point Observatory 3.5-meter telescope, which is owned and operated by the Astrophysical Research Consortium. S.D. Benecchi acknowledges funding support from DTM's Carnegie Fellowship and the use of telescopes at Las Campanas Observatory for this work. A.J. Verbiscer acknowledges support from NASA Planetary Astronomy grant NNX09AC99G. We thank David Tholen and an anonymous reviewer for making constructive suggestions that substantially improved this paper. Finally, we thank the free and open source software communities for empowering us with key tools used to complete this project, notably Linux, the GNU tools, LibreOffice, MySQL, Evolution, Python, and FVWM.